\documentclass[twocolumn,english,aps,prl,twocolumn,superscriptaddress,floatfix]{revtex4-1}
\usepackage[T1]{fontenc}
\usepackage[latin9]{inputenc}
\usepackage{color}
\usepackage{babel}
\usepackage{amsmath}
\usepackage{amssymb}
\usepackage{graphicx}
\usepackage[unicode=true,pdfusetitle,
 bookmarks=false,
 breaklinks=false,pdfborder={0 0 1},backref=false,colorlinks=false]
 {hyperref}
\hypersetup{
 colorlinks,linkcolor=blue,citecolor=blue,urlcolor=blue}

\makeatletter
\usepackage{babel}

\makeatother

\begin{document}

\title{Microscopic Theory of Spin Relaxation Anisotropy in Graphene with
Proximity-Induced Spin--Orbit Coupling}

\author{Manuel Offidani }

\affiliation{University of York, Department of Physics, YO10 5DD, York, United
Kingdom}

\author{Aires Ferreira}

\affiliation{University of York, Department of Physics, YO10 5DD, York, United
Kingdom}
\begin{abstract}
Inducing sizable spin--orbit interactions in graphene by proximity
effect is establishing as a successful route to harnessing two-dimensional
Dirac fermions for spintronics. Semiconducting transition metal dichalcogenides
(TMDs) are an ideal complement to graphene because of their strong
intrinsic spin--orbit coupling (SOC) and spin/valley-selective light
absorption, which allows all-optical spin injection into graphene.
In this study, we present a microscopic theory of spin dynamics in
weakly disordered graphene samples subject to uniform proximity-induced
SOC as realized in graphene/TMD bilayers. A time-dependent perturbative
treatment is employed to derive spin Bloch equations governing the
spin dynamics at high electronic density. Various scenarios are predicted,
depending on a delicate competition between interface-induced Bychkov-Rashba
and spin--valley (Zeeman-type) interactions and the ratio of intra-
to inter-valley scattering rates. For weak SOC compared to the disorder-induced
quasiparticle broadening, the anisotropy ratio of out-of-plane to
in-plane spin lifetimes $\zeta=\tau_{s}^{\perp}/\tau_{s}^{\parallel}$
agrees qualitatively with a toy model of spins in a weak fluctuating
SOC field recently proposed by Cummings\emph{ }and co-workers {[}PRL
\textbf{119}, 206601 (2017){]}. In the opposite regime of well-resolved
SOC, qualitatively different formulae are obtained, which can be tested
in ultra-clean heterostructures characterized by uniform proximity-induced
SOC in the graphene layer. 
\end{abstract}
\maketitle

\section{INTRODUCTION}

The tailored control of electronic properties in van der Waals heterostructures
built from the assembly of two-dimensional (2D) crystals has provided
a unique route to explore interface-induced phenomena \citep{Belleste_Nanoscale2011,Butler_ACSNano2013,Zhang_JPDAppl2017}.
Heterostructures combining graphene and semiconducting group-VI dichalcogenides
{[}MX$_{2}$ (e.g., M$=$Mo, W; X$=$S, Se){]} could enable low-power
spin-logic devices harnessing the unique interplay between quantum
(spin and valley) degrees of freedom in honeycomb layers \citep{Zibouche14,Soumyanarayanan_16,Garcia18}.
This thrust has been fueled by the prospect of enhancing spin--orbital
effects in graphene \citep{Geim_Nature2013,Soumyanarayanan_Nature2016},
while preserving the quintessential Dirac character of its 2D quasiparticles.
The much sought after interface-induced SOC has been recently demonstrated
in graphene/TMD bilayer heterostructures \citep{Avsar_NatComm2014,Wang_NatComm2015,Wang_PRX2016,Yang_2DMat2016,Volk_PRB2017,Zihlmann_PRB2018},
where sharp weak antilocalization features in the magnetoconductance
data \citep{Wang_PRX2016,Yang_2DMat2016,Volk_PRB2017,Zihlmann_PRB2018}
and dramatic reduction of spin lifetimes \citep{Raes_NatComm2016,Benitez_Nature2017,Ghiasi_NanoLett2017}
hint at a massive enhancement of spin--orbit interactions in the
2D carbon layer (up to 10 meV), consistent with the predictions of
model calculations and first-principles studies \citep{Wang_NatComm2015,Gmitra_PRB2016,Alsharari_PRBR2016}. 

The modification of electronic states in graphene-based van der Waals
heterostructures due to proximity-induced SOC can be understood within
a weak interlayer coupling picture, where Dirac states located in
the band gap of a 2D semiconductor are perturbed in two \textcolor{black}{fundamental
ways. }Firstly, the interfacial breaking of mirror inversion symmetry
leads to the familiar Bychkov-Rashba effect \citep{BychkovRashba84}.
The spin rotational invariance is lifted (point group symmetry reduction
$D_{6h}\rightarrow C_{6v}$), which causes the spin splitting of the
Dirac states. Secondly, the proximity to different atoms (metal or
chalcogen elements) located beneath the graphene flake ($C_{6v}\rightarrow C_{3v}$)
effectively ``transfers'' the sublattice-resolved SOC of the TMD
substrate onto graphene (and hence spin--valley interactions). The
relative magnitude of the spin--orbit effects experienced by $\pi$-electrons
in graphene depend on type and number of TMD layers, degree of vertical
strain, and possible presence of resonant spin--orbit scatterers
\citep{Gmitra16,Singh_GTMDs2018,Pachoud15,Huang16,Wakamura18}. The
proximity spin--orbital effects couple all internal degrees of freedom
of graphene (i.e. spin, sublattice and valley), enabling interesting
spin-dependent non-equilibrium phenomena, including highly anisotropic
spin dynamics \citep{Cummings_PRL2017}, spin-Galvanic and spin-Hall
effects \citep{Offidani_PRL2017,Milletari_PRL2017,Garcia_NanoLett2017}. 

In this work, we investigate how spin relaxation times in weakly disordered
monolayer graphene are affected by proximity-induced SOC. The spin--orbit
(SO) interaction enters the long-wavelength continuum Hamiltonian
as an additional uniform term $V_{\text{SO}}$, that is (we choose
natural units with $\hbar=1=e$)
\begin{equation}
H_{C_{3v}}=\int d\mathbf{x}\,\Psi^{\dagger}(\mathbf{x})\left[\tau_{z}\,v\,\boldsymbol{\sigma}\cdot\mathbf{p}+V_{\textrm{SO}}+U(\mathbf{x})\,\right]\Psi(\mathbf{x})\,,\label{eq:1}
\end{equation}
where $v$ is the Fermi velocity of massless Dirac fermions and $U(x)$
is a disorder potential describing scattering from nonmagnetic impurities.
The Hamiltonian is expressed in the basis $(KA\uparrow,KA\downarrow,KB\uparrow,KB\downarrow,K^{\prime}B\uparrow,K^{\prime}B\downarrow,K^{\prime}A\uparrow,K^{\prime}A\downarrow)^{T}$
and we have introduced $\tau_{\varsigma}$ ($\sigma_{\varsigma}$)
with $\varsigma=0,x,y,z$ as Pauli matrices in the valley (sublattice)
space, respectively (here, $\tau_{0}$ and $\sigma_{0}$ denote identity
matrices). While knowing exactly the SO interaction is generally not
possible, first-principles calculations and transport data provide
a mean to estimate the various SO terms allowed by symmetry \citep{Bir_Book1974,Winkler_Book2003,Kochan_PRB2017,Cysne_PRB2018}.
It is straightforward to show that there are only three such terms
compatible with $C_{3v}$ symmetry, $V_{\textrm{SO}}=H_{\text{KM}}+H_{\text{BR}}+H_{\text{sv}}$,
respectively, intrinsic-like SOC \citep{McClure_1962,Kane_2005},
Bychkov-Rashba SOC \citep{Bychov_1984} and spin--valley interaction
\citep{Gmitra_PRB2016,Kochan_PRB2017}. We note in passing that, beyond
SOC, charge carriers in graphene can also experience an orbital sublattice-staggered
potential $H_{\Delta}=\Delta\,\tau_{z}\sigma_{z}$ \citep{Gmitra_PRB2016}.
This effect is believed to be very weak in graphene/TMD bilayers (in
contrast to rotationally aligned graphene on h-BN \citep{Jung_NanoLett2012})
and will be neglected in the following discussion \citep{Foot_1}.

The intrinsic-type SOC reads 
\begin{equation}
H_{\text{KM}}=\lambda_{\text{KM}}\,\tau_{0}\,\sigma_{z}\,s_{z}.\label{eq:KM_Ham}
\end{equation}
where $\lambda_{\text{KM}}$ is the spin--orbit energy. This term
is invariant under all symmetry operations of the $D_{6h}$ group,
and thus it is already present in pristine graphene. As shown in a
seminal work by Kane and Mele \citep{Kane_2005}, a large $\lambda_{\text{KM}}$
would drive graphene into a nontrivial $\mathbb{Z}_{2}$ topological
insulating phase. However this term is very weak in graphene on typical
substrates \citep{Min_PRB2006,Hernando_PRB2006,Gmitra_PRB2009}. Furthermore,
in 2D heterostructures, the interfacial breaking of mirror inversion
symmetry favours the appearance of an in-plane pseudo-magnetic field,
that is, the familiar Bychkov-Rashba effect. This term (invariant
under the $C_{6v}$ point group) directly couples to the velocity
of electrons, thus acting as a Lorentz pseudomagnetic field \citep{Milletari_PRL2017}:
\begin{equation}
H_{\text{BR}}=\lambda\,\tau_{z}\,\hat{z}\cdot(\boldsymbol{\sigma}\times\mathbf{s})\,.\label{eq:Rashba_SOC}
\end{equation}
Finally, in honeycomb layers with interpenetrating triangular lattices
made up of chemically distinct species, another spin-conserving SOC
is allowed \citep{Zhu_PRB2011,Xu_NatPhys2014}. The sublattice inversion
asymmetry can be captured by introducing sublattice-resolved next-nearest
neighbours hoppings reducing the point group symmetry to $C_{3v}$
\citep{Kochan_PRB2017}. This leads to a Zeeman-type spin-valley coupling
\begin{equation}
H_{\text{sv}}=\lambda_{\text{sv}}\,\tau_{z}\,\sigma_{0}\,s_{z}\,.\label{eq:SV_Ham}
\end{equation}
The $C_{3v}$ scenario faithfully describes graphene on TMDs, where
the small lattice mismatch produces different SO energy on $A,B$
carbon sublattices \citep{Wang_NatComm2015,Kochan_PRB2017,Gmitra_PRB2016}.
Uniform proximity-induced SO terms are block diagonal in valley space
due to absence of interlayer hoppings connecting inequivalent valleys
in graphene \citep{Phong_2Dmat2017,Cysne_PRB2018}. 
\begin{figure}
\centering\includegraphics[width=1\columnwidth]{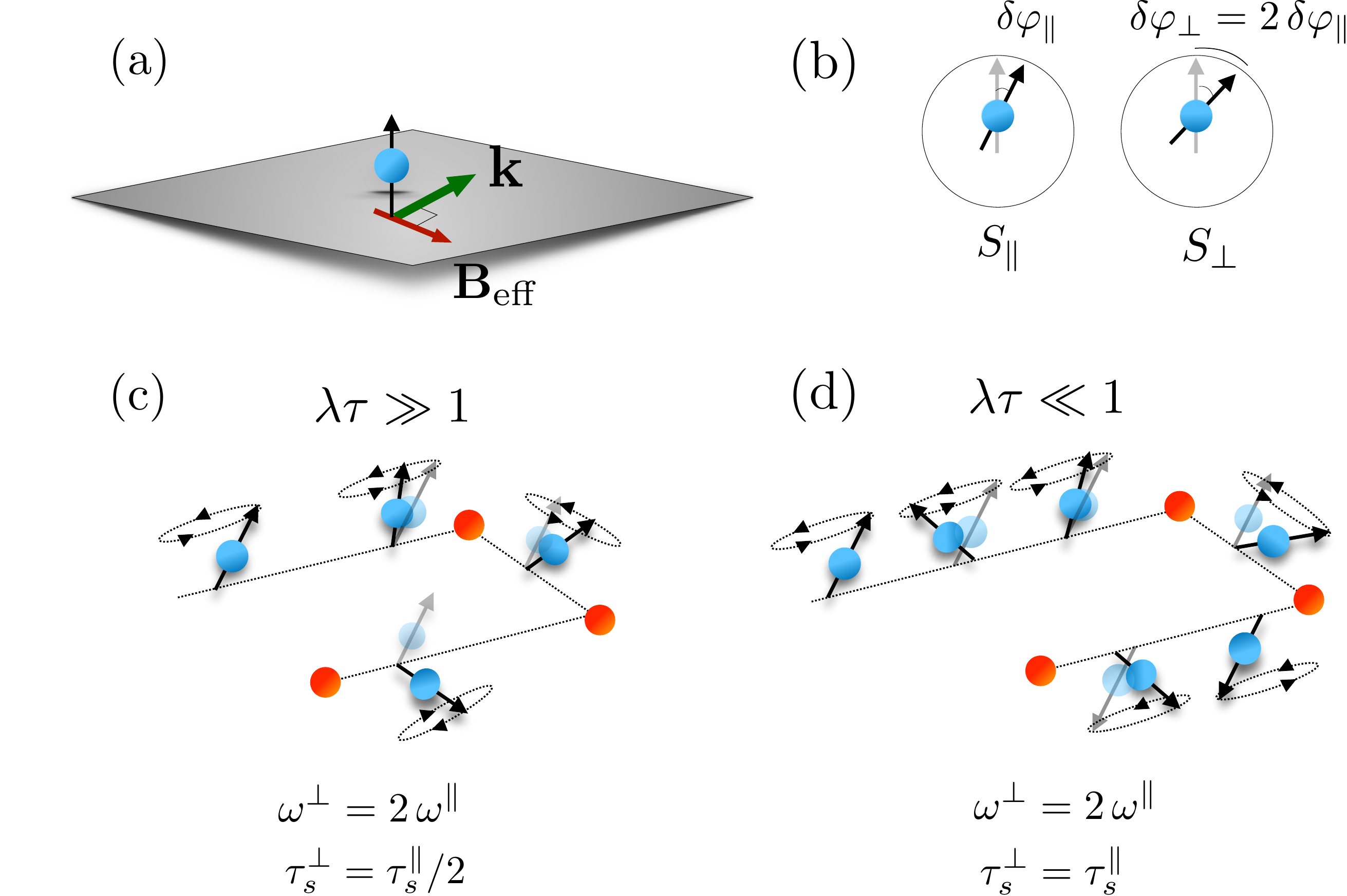}\caption{Spin relaxation in the minimal $C_{6v}$ model with Bychkov-Rashba
interaction. (a) The SO field $\mathbf{B}_{\text{eff}}$ is orthogonal
to the direction of motion $\mathbf{\hat{k}}$. (b) Due to the in-plane
character of the Bychkov-Rashba interaction, initially $\hat{z}$-polarized
spins $S_{\perp}$ are subject to a precession twice as fast as in-plane
$S_{\parallel}$ spins. This reflects in: (c) a twice shorter spin
relaxation time $\tau_{s}^{\perp}=\tau_{s}^{\parallel}/2$ when $\lambda\tau\ll1$
; (d) a faster precession period in the damped oscillating mode (see
Eqs.\,(\ref{eq:weakRashba_Sz_intra}), (\ref{eq:Sx_Rashba})), with
a isotropic spin relaxation time $\tau_{s}^{\perp}=\tau_{s}^{\parallel}$,
when $\lambda\tau\gg1$. \label{fig:Spin-relaxation-in} }
\end{figure}

Interface-induced Bychkov-Rashba and spin-valley interactions in graphene/TMD
bilayers can in principle be large as tens of meV. With such a sizable
imprinted in-plane (Lorentz-type) and out-of-plane (Zeeman-type) SO
fields, the spin relaxation times for in-plane ($\parallel$) and
out-of-the-plane ($\perp$) polarization channels can be dramatically
different. A recently-introduced figure of merit for the competition
of the SOC along orthogonal spatial directions is the \emph{spin relaxation
time anisotropy} (SRTA): $\zeta=\tau_{\perp}/\tau_{\parallel}$, which
in graphene on TMDs has been estimated to be of order $\zeta\sim10-100$
\citep{Raes_NatComm2016,Benitez_Nature2017,Ghiasi_NanoLett2017}.
A simple treatment to obtain SRTA ratios has been put forward in Ref.\,\citep{Fabian_Acta2007},
which assumes that the electronic motion of bare quasiparticles (without
SOC) is affected by a perturbing spin--orbit field with its precession
axis randomly changing due to impurity scattering. The model applied
to graphene/TMDs systems yields analytic formulas relating $\zeta$
to the ratio $\lambda_{\text{sv}}/\lambda$ and $\tau/\text{\ensuremath{\tau_{\text{iv}}}}$,
where $\tau$ and $\tau_{\text{iv}}$ are, respectively, the intra-
and inter-valley momentum lifetimes \citep{Cummings_PRL2017}. However,
the formalism presented there is limited to weak SOC, that is, $\lambda_{\text{SOC}}\tau\ll1$
with $\lambda_{\text{SOC}}=\{\lambda,\lambda_{\textrm{sv}}\}$. This
can be a strong constraint when trying to model ultra-clean samples
with high charge carrier mobility, in which $\lambda_{\text{SOC}}\tau$
can be as large as unity \citep{Wang_PRX2016}. Also, a \emph{microscopic}
approach able to provide more physical insight to spin relaxation
would be desirable, calling for a detailed study of how the spin dynamics
is affected by the interplay of uniform proximity-induced SOC and
impurity scattering. Here, we address theoretically this problem by
means of the single-particle density matrix formalism. We obtain a
set of coupled spin Bloch equations governing the spin dynamics for
high electronic density $\epsilon\gg\lambda,\,\lambda_{\text{sv}}$---$\epsilon$
being the Fermi energy---assuming Gaussian-type (white-noise) disorder
leading to intra- and inter-valley scattering processes. A variety
of scenarios is shown to emerge, from simple- or multi-exponentially
decaying spin dynamics to purely damped oscillating modes, depending
on the relative magnitude of the three main energy scales: $\lambda$,
$\lambda_{\text{sv}}$ and $1/\tau$. We provide analytic expressions
for the SRTA $\zeta$ in the asymptotic limits of weak SOC (compatible
with the findings of Ref.\,\citep{Cummings_PRL2017}) and strong
SOC, which should be used to fit experimental data when $\lambda_{\text{SOC}}\tau\gtrsim1$.

The paper is organized as follows. Section\,\,II derives the general
spin Bloch equations starting from the quantum Liouville equation.
In Sec.\,III, we provide analytic solutions in the presence or absence
of intervalley scattering and in the limiting cases of weak and strong
SOC. Section\,\,IV discusses the obtained SRTA, putting it in relation
with recent theoretical and experimental results and Sec.\,IV presents
our conclusions.

\section{Formalism: Spin bloch equations}

The starting point of our approach is the quantum Liouville equation
for the single-particle density matrix operator \citep{Tarasenko06,Culcer_PRB2007_2008,Dugaev09,Wu_2010}
\begin{align}
\frac{\partial\rho}{\partial t} & =\imath[\,H_{0}+V_{\text{SO}}+U,\rho\,]\,.\label{eq:Quantum_Liouville}
\end{align}
We consider a scattering potential $U$ generated by dilute short-range
impurities at random locations $\{\mathbf{x}_{i}\}_{i=1...N}$, 
\begin{equation}
U(\mathbf{x})=U_{\textrm{intra}}(\mathbf{x})+U_{\textrm{inter}}(\mathbf{x})=\sum_{i=1}^{N}(u_{i}+w_{i}\tau_{x})f_{i}(\mathbf{x})\,,\label{eq:U_scattering}
\end{equation}
where $u_{i}$ ($w_{i}$) are reals parameterizing the amplitude of
intravalley (intervalley) scattering processes and $\{f_{i}(\mathbf{x})\}$
characterize the spatial profile of the scattering potential. In order
to derive the spin Bloch equations for high electronic density, we
follow closely the treatment by Culcer and Winkler \citep{Culcer_PRB2007_2008}.
The first step is to project Eq.\,(\ref{eq:Quantum_Liouville}) onto
plane-wave eigenstates of the unperturbed graphene Hamiltonian, namely
\begin{equation}
|\mathbf{k}\sigma\kappa s\rangle=\frac{1}{\sqrt{2}}\left(\begin{array}{c}
\kappa\,\sigma\,e^{-\imath\phi_{\mathbf{k}}/2}\\
e^{\imath\phi_{\mathbf{k}}/2}
\end{array}\right)\otimes|\kappa\rangle\otimes|s\rangle\,,\label{eq:Bloch_eigenstate}
\end{equation}
where $\mathbf{k}$ is the wavevector around a Dirac point ($\phi_{\mathbf{k}}$
is the wavevector angle) and $\sigma,\kappa,s=\pm1$ are quantum indices
for sublattice, valley and spin, respectively. The free eigenvalues
read as $\epsilon_{\mathbf{k}}^{\sigma\kappa s}=\sigma vk$, where
$k=|\mathbf{k}|$. $\rho$ is then a matrix of dimension $2^{3}=8$,
whose matrix elements are written as $\rho_{\mathbf{kk^{\prime}}}\equiv\rho_{\mathbf{kk}^{\prime}}^{\alpha\alpha^{\prime}}=\langle\mathbf{k^{\prime}\alpha^{\prime}|\rho}|\mathbf{k}\alpha\rangle$
and $\alpha=\{\sigma,\kappa,s\}$ is short-hand for the set of quantum
indices (we use a similar notation for $H_{0}$, $V_{\text{SO}}$
and $U$). The proximity-induced SOC term $V_{\text{SO}}$ has non-zero
matrix elements between conduction and valence states leading to interband
transitions. However, we focus here on the large Fermi energy regime
$\epsilon/\lambda_{\text{SOC}}\gg1$, where interband coherence effects
are strongly suppressed. Hence, we take $\langle\mathbf{k}^{\prime}\sigma^{\prime}|\rho|\mathbf{k}\sigma\rangle=\delta_{\sigma\sigma^{\prime}}\rho_{\mathbf{kk^{\prime}}}^ {}$.
For simplicity of notation, we consider positive energies $\epsilon>0$,
henceforth considering electrons in the conduction band $\sigma=1$
and dropping the sublattice index from all expressions. To simplify
the treatment we also neglect valley coherence $\langle\kappa^{\prime}|\rho|\kappa\rangle=\delta_{\kappa\kappa^{\prime}}$
\citep{Foot_2}. The two inequivalent Dirac points $K,K^{\prime}$
can only be connected then by scattering events, according to Eq.\,(\ref{eq:U_scattering}).

Following Ref.\,\citep{Culcer_PRB2007_2008}, we split the density
matrix into diagonal and off-diagonal elements: $\rho_{\mathbf{kk^{\prime}}}=f_{\mathbf{k}}\delta_{\mathbf{kk^{\prime}}}+g_{\mathbf{kk^{\prime}}}\,,$where
for $g_{\mathbf{kk^{\prime}}}$ it is assumed $\mathbf{k}\neq\mathbf{k}^{\prime}$.
We have 
\begin{align}
\frac{df_{\mathbf{k}}}{dt}+\imath[H_{0}+V_{\text{SO}},f_{\mathbf{k}}] & =-\imath[U,g]_{\mathbf{kk}}\,,\label{eq:eq_f}\\
\frac{dg_{\mathbf{kk^{\prime}}}}{dt}+\imath[H_{0},g]_{\mathbf{kk^{\prime}}} & =-\imath[U,g]_{\mathbf{kk^{\prime}}}\,.\label{eq:eq_g}
\end{align}
To simplify the analytical treatment, we neglect the term $V_{\text{SO}}$
in the commutator on the left-hand side of Eq.\,(\ref{eq:eq_g}).
The approximation is valid in the limit of high Fermi energy, that
is, $\epsilon\gg\lambda_{\text{SOC}}$. Also, $U$ only contains off-diagonal
elements in $\mathbf{k}$, such that the commutator on the right-hand
side of Eq.\,\eqref{eq:eq_f} only contains $g$. We are ultimately
interested in the diagonal part $f$, as the spin observables are
defined as 
\begin{equation}
\mathbf{S}=\frac{1}{2}\text{Tr}[\rho\cdot\mathbf{s}]=\frac{1}{2}\sum_{\mathbf{k},\kappa}\text{tr}[f_{\mathbf{k}}^{\kappa}\cdot\mathbf{s}]=\frac{1}{2}\sum_{\mathbf{k},\kappa}\mathbf{S}_{\mathbf{k}}^{\kappa}\,.\label{eq:Spin_Observables}
\end{equation}
We hence solve Eq.\,(\ref{eq:eq_g}) and substitute the solution
into the right-hand side of Eq.\,(\ref{eq:eq_f}), which gives the
collision integral. As customary, we treat Eq.\,(\ref{eq:eq_g})
perturbatively for weak disorder with Gaussian (white-noise) statistics
\begin{align}
\langle U_{\mathbf{kk^{\prime}}}^{\alpha\alpha^{\prime}}\rangle_{\text{dis}} & =0\,,\label{eq:u_1}\\
\langle U_{\mathbf{kk^{\prime}}}^{\alpha\alpha^{\prime}}U_{\mathbf{k^{\prime}k^{\prime\prime}}}^{\alpha^{\prime}\alpha^{\prime\prime}}\rangle_{\text{dis}} & =\delta_{\mathbf{k},\mathbf{k}^{\prime\prime}}\text{\ensuremath{\delta}}_{\alpha\alpha^{\prime\prime}}\,n_{i}\,|U_{\mathbf{kk^{\prime}}}^{\alpha\alpha^{\prime}}|^{2}\,,\label{eq:u_2}
\end{align}
where $n_{i}$ is the impurity areal density. After a somewhat lengthy
but straightforward calculation, where Eqs.\,\eqref{eq:eq_f}-\eqref{eq:eq_g}
are expressed in the interaction picture and the evolution operator
 is expanded in powers of $U$, one arrives at the following equation
for the spin components $\mathbf{S}_{\mathbf{k}}^{\kappa}$ \begin{widetext}

\begin{align}
\partial_{t}\mathbf{S}_{\mathbf{k}}^{\kappa}+\imath\,\mathbf{L}_{\mathbf{k}}^{\kappa}\cdot\mathbf{S}_{\mathbf{k}}^{\kappa} & =-\pi\sum_{\mathbf{k^{\prime}}\kappa^{\prime}}\delta(\epsilon_{\mathbf{k}}-\epsilon_{\mathbf{k}^{\prime}})\,\langle\mathbf{S}_{\mathbf{k}}^{\kappa}\,U_{\mathbf{kk^{\prime}}}^{\kappa\kappa^{\prime}}\,U_{\mathbf{k}^{\prime}\mathbf{k}}^{\kappa^{\prime}\kappa}+U_{\mathbf{kk^{\prime}}}^{\kappa\kappa^{\prime}}\,U_{\mathbf{k}^{\prime}\mathbf{k}}^{\kappa^{\prime}\kappa}\,\mathbf{S}_{\mathbf{k}}-2\,U_{\mathbf{kk^{\prime}}}^{\kappa\kappa^{\prime}}\,\mathbf{S}_{\mathbf{k^{\prime}}}^{\kappa^{\prime}}\,U_{\mathbf{k}^{\prime}\mathbf{k}}^{\kappa^{\prime}\kappa}\rangle_{\text{dis}}\,\label{eq:Spin_Bloch}
\end{align}
with a Larmor precession term 
\begin{equation}
\mathbf{L}_{\mathbf{k}}^{\kappa}=\left(\begin{array}{ccc}
0 & -\kappa\lambda_{\text{sv}} & \lambda\cos\phi_{\mathbf{k}}\\
\kappa\lambda_{\text{sv}} & 0 & \lambda\sin\phi_{\mathbf{k}}\\
-\lambda\cos\phi_{\mathbf{k}} & -\lambda\sin\phi_{\mathbf{k}} & 0
\end{array}\right)\,.\label{eq:Larmor_Prec}
\end{equation}
\end{widetext}A few comments are in order. Central to the derivation
of the quantum kinetic equation for the reduced spin density matrix
{[}Eq.~(\ref{eq:Spin_Bloch}){]} is the assumption of Gaussian disorder.
The latter is equivalent to the first Born approximation \citep{Offidani_MDPI2018}
and thus it neglects any effects from skew scattering (allowed in
the $C_{3v}$ model \citep{Milletari_PRL2017}) and modifications
to the energy dependence of the collision integral due to scattering
resonances. Nevertheless, the relation between spin lifetime and momentum
scattering time is expected to be preserved at all orders in perturbation
theory, as shown explicitly in the minimal Dirac--Rashba model ($\lambda_{\textrm{sv}}=0$)
with $\lambda_{\text{SOC}}\tau\ll1$ \citep{Offidani_MDPI2018}. This
means that inclusion of higher-order scattering processes beyond the
first Born approximation should not affect the SRTA ratios in the
regime of validity of the quantum kinetic treatment ($\epsilon\tau\gg1$),
consistently with the findings from exact numerical simulations \citep{Cummings_PRL2017}. 

Next, we use the quantum kinetic equation Eq.\,(\ref{eq:Spin_Bloch})
to obtain the spin Bloch equations governing the spin dynamics. Firstly,
we separate the collision integral $I[\mathbf{S}_{\mathbf{k}}^{\kappa}]$
into intra and inter-valley parts, $\kappa^{\prime}=\{\kappa,\bar{\kappa}\}=\{\kappa,-\kappa\}$,
with the corresponding matrix elements of the scattering potential
\begin{align}
|U_{\mathbf{kk^{\prime}}}^{\kappa\kappa}|^{2} & =u^{2}\cos^{2}\frac{\phi_{\mathbf{k}}-\phi_{\mathbf{k^{\prime}}}}{2}\equiv u^{2}F_{\mathbf{kk^{\prime}}}\,,\\
|U_{\mathbf{kk^{\prime}}}^{\kappa\bar{\kappa}}|^{2} & =w^{2}\sin^{2}\frac{\phi_{\mathbf{k}}-\phi_{\mathbf{k^{\prime}}}}{2}\equiv w^{2}G_{\mathbf{kk^{\prime}}}\,,
\end{align}
where we have assumed that the impurity potential has a common matrix
structure i.e., $u_{i}=u$ and $w_{i}=w$ (the generalization of our
results to an arbitrary number of uncorrelated disorders can be easily
accomplished using the standard Mathiessen's rule). We can then write
\begin{align}
I^{\text{intra}}[\mathbf{S}_{\mathbf{k}}^{\kappa}] & =-2\pi\,n_{i}u^{2}\sum_{\mathbf{k^{\prime}}}F_{\mathbf{kk^{\prime}}}(\mathbf{S}_{\mathbf{k}}^{\kappa}-\mathbf{S}_{\mathbf{k^{\prime}}}^{\kappa})\,\Delta_{\mathbf{k}\mathbf{k}^{\prime}}\,,\label{eq:I_intra}\\
I^{\text{inter}}[\mathbf{S}_{\mathbf{k}}^{\kappa}] & =-2\pi\,n_{i}w^{2}\sum_{\mathbf{k^{\prime}}}G_{\mathbf{kk^{\prime}}}(\mathbf{S}_{\mathbf{k}}^{\kappa}-\mathbf{S}_{\mathbf{k^{\prime}}}^{\bar{\kappa}})\,\Delta_{\mathbf{k}\mathbf{k}^{\prime}}\,.\label{eq:I_inter}
\end{align}
where $\Delta_{\mathbf{k}\mathbf{k}^{\prime}}\equiv\delta(\epsilon_{\mathbf{k}}-\epsilon_{\mathbf{k}^{\prime}})$.
To solve the coupled system of 6 equations (3 spin $\times$ 2 valley)
Eq.\,(\ref{eq:Spin_Bloch}), we expand $\mathbf{S_{k}^{\kappa}}$
in cylindric harmonics

\begin{equation}
\mathbf{S}_{\mathbf{k}}^{\kappa}=\sum_{m=-\infty}^{\infty}\mathbf{S}_{k}^{\kappa,m}e^{\imath\,m\,\phi_{\mathbf{k}}}\,,\label{eq:S_harmonics}
\end{equation}
We note that the Dirac-delta function in Eq.\,(\ref{eq:Spin_Bloch})
imposes energy conservation i.e., $k=k^{\prime}$, such that the components
of $\mathbf{S}_{\mathbf{k}^{\prime}}^{\kappa}$ also depend on $k$.
Substituting Eq.\,(\ref{eq:S_harmonics}) into Eq.\,(\ref{eq:Spin_Bloch}),
and retaining only the lowest-order harmonics $m=0,\pm1$ we finally
obtain (see Appendix for details) 
\begin{align}
\partial_{t}S_{x}^{0} & =-\frac{2\alpha^{2}}{\tau}(S_{x}^{0}-\bar{S}_{x}^{0})-2\lambda_{\text{sv}}S_{y}^{0}+\lambda\sum_{m=\pm1}S_{z}^{m}\,,\label{eq:system_11}\\
\partial_{t}S_{y}^{0} & =-\frac{2\alpha^{2}}{\tau}(S_{y}^{0}-\bar{S}_{y}^{0})+2\lambda_{\text{sv}}S_{x}^{0}+\imath\lambda\sum_{m=\pm1}mS_{z}^{m}\,,\label{eq:system_12}\\
\partial_{t}S_{z}^{0} & =-\frac{2\alpha^{2}}{\tau}(S_{z}^{0}-\bar{S}_{z}^{0})-\lambda\sum_{m=\pm1}\left(S_{x}^{m}+\imath mS_{y}^{m}\right)\,,\label{eq:system_13}
\end{align}
and 
\begin{align}
\partial_{t}S_{x}^{\pm1} & =\lambda S_{z}^{0}-2\lambda_{\text{sv}}S_{y}^{\pm1}-h_{\alpha}(S_{x}^{\pm},\bar{S}_{x}^{\pm})\,,\label{eq:system_21}\\
\partial_{t}S_{y}^{\pm1} & =\mp\imath\lambda S_{z}^{0}+2\lambda_{\text{sv}}S_{x}^{\pm1}-h_{\alpha}(S_{y}^{\pm},\bar{S}_{y}^{\pm})\,,\label{eq:system_22}\\
\partial_{t}S_{z}^{\pm1} & =-\lambda(S_{x}^{0}\mp\imath S_{y}^{0})-h_{\alpha}(S_{z}^{\pm},\bar{S}_{z}^{\pm})\,,\label{eq:system_23}
\end{align}
where 
\begin{equation}
h_{\alpha}(S_{i}^{\pm1},\bar{S}_{i}^{\pm1})=\frac{1}{\tau}[(1+2\alpha^{2})S_{i}^{\pm1}+\alpha^{2}\bar{S}_{i}^{\pm1}]\,,
\end{equation}
with $(\mathbf{S}_{k}^{\kappa=\pm1,m})_{i}\equiv(S_{i}^{m},\bar{S}_{i}^{m})$.
We have introduced the ratio of inter- to intra-valley energy scales
defined as\textbf{ $\alpha=w/u$}, as well as the intravalley momentum
scattering time\textbf{ 
\begin{equation}
\tau=(n_{i}u^{2}\epsilon/4v^{2})^{-1}\,.\label{eq:intravalley}
\end{equation}
}The spin Bloch equations {[}Eqs.~(\ref{eq:system_11})-(\ref{eq:system_23}){]}\textcolor{black}{{}
together with the corresponding expressions for the barred component
at $\kappa=-1$---obtained by the formal replacement $S\to\bar{S}$
and $\lambda_{\text{sv}}\to-\lambda_{\text{sv}}$---}are the central
result of this section.\textcolor{black}{{} }

\section{Results}

\textcolor{black}{We are mostly interested in the zeroth harmonics
of the various spin components, which according to Eq.\,(\ref{eq:Spin_Observables})
completely determine the spin density observables \citep{foot_3}.
In most cases it is not possible to derive a simple closed expressions
for arbitrary $\lambda,\lambda_{\text{sv}}$. Therefore in the following
we solve the equations in the two limiting cases $\lambda\gg\lambda_{\text{sv}}$
and $\lambda\ll\lambda_{\text{sv}}$, which is also helpful to get
physical insight.}

\subsection{Intravalley scattering only: $w=0$}

The calculations are carried out explicitly for the out-of-plane component
$S_{z}\equiv S_{z}^{0}+\bar{S}_{z}^{0}$. The spin Bloch equations
are recast in the following form 
\begin{equation}
\left(\begin{array}{ccc}
\partial_{t} & -\lambda & 0\\
4\lambda & \partial_{t}-\frac{1}{\tau} & -2\lambda_{\text{sv}}\\
0 & 2\lambda_{\text{sv}} & \partial_{t}-\frac{1}{\tau}
\end{array}\right)\left(\begin{array}{c}
S_{z}\\
y\\
z
\end{array}\right)=\left(\begin{array}{c}
0\\
0\\
0
\end{array}\right)\,,\label{eq:Sz0}
\end{equation}
where we introduced the following admixtures of in-plane spin harmonics
\begin{align}
y & =\sum_{m=\pm1}(S{}_{x}^{m}+\bar{S}_{x}^{m})+\imath\,m\,(S_{y}^{m}+\bar{S}_{y}^{m})\,,\label{eq:y}\\
z & =\sum_{m=\pm1}(S_{y}^{m}+\bar{S}_{y}^{m})-\imath\,m\,(S_{x}^{m}+\bar{S}_{x}^{m})\,.\label{eq:z}
\end{align}
The eigenfunctions can be written as 
\begin{equation}
\left(\begin{array}{c}
S_{z}(t)\\
y(t)\\
z(t)
\end{array}\right)=\sum_{i=1}^{3}c_{i}\mathbf{v}_{i}e^{\omega_{i}t}\,,
\end{equation}
where 
\begin{figure}
\centering{}\includegraphics[width=0.75\columnwidth]{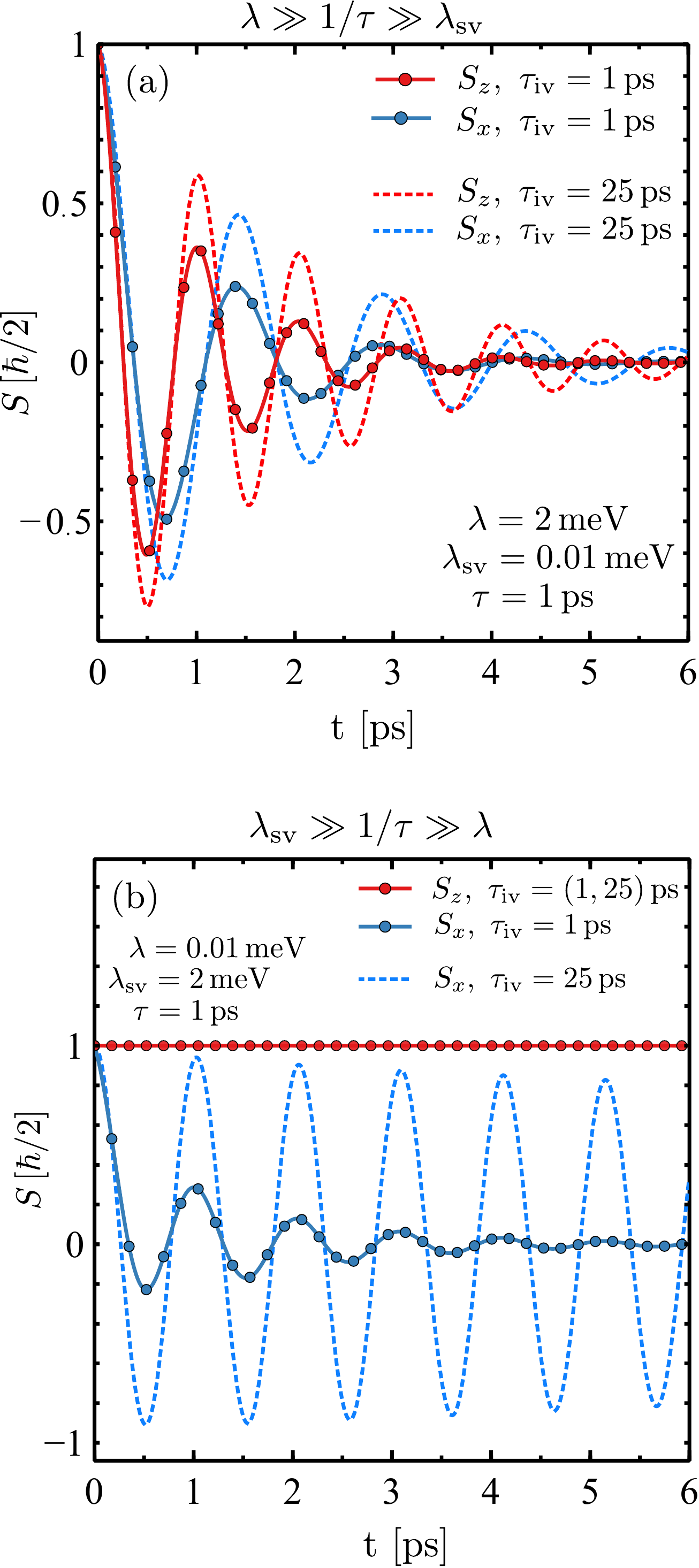} \caption{Spin dynamics for strong proximity-induced SOC ($\lambda\tau,\lambda_{\textrm{sv}}\tau\gg1$)
in the presence of intervalley scattering. For dominant Bychkov-Rashba
SOC (a), while the period of the oscillation is different for in-plane
and out-of-plane spins, the decaying (spin relaxation) time is the
same, as discussed in the main text and illustrated in Fig.\,(\ref{fig:Spin-relaxation-in}).
For dominant spin-valley SOC (b) the out-of-plane component is\textbf{
}weakly sensitive to the value of $\tau_{\text{iv}}$. This is expected
to hold in the highly-doped regime $\epsilon\gg\lambda_{\text{sv}}$,
as discussed in the main text. \label{fig:Spin-dynamics-for}}
\end{figure}
$\omega_{i}$ are the the solution of the algebraic equation 
\begin{equation}
\omega^{3}+\frac{2}{\tau}\omega^{2}+\left[4(\lambda^{2}+\lambda_{\text{sv}}^{2})+\frac{1}{\tau^{2}}\right]\omega+\frac{4\lambda^{2}}{\tau}=0\,,\label{eq:omega_Zeta}
\end{equation}
and $\mathbf{v}_{i}$ are the corresponding eigenvectors. The coefficients
$c_{i}$ are determined by imposing the Cauchy boundary conditions
$S_{z}^{0}(t=0)=1,\,y(t=0)=z(t=0)=0$. The analytical solution to
Eq.\,(\ref{eq:omega_Zeta}) is rather cumbersome. It is more transparent
instead to find a solution perturbatively by expanding 
\begin{equation}
\omega=\omega^{(0)}+\beta\,\omega^{(1)}+\beta^{2}\omega^{(2)}+O(\beta^{3})\,,
\end{equation}
where $\beta\ll1$ and $\beta=\lambda_{\text{sv}}/\lambda$ ($\beta=\lambda/\lambda_{\text{sv}}$)
representing the case of dominant Bychkov-Rashba (spin-valley) spin--orbit
interaction. We find for $\lambda\gg\lambda_{\text{sv}}$ 
\begin{align}
S_{z}(t) & =\sum_{s=\pm1}\frac{1}{2}(1-\frac{s}{\sqrt{1-c_{z}^{2}}})e^{\omega_{s}t}\,,
\end{align}
where $c_{z}=4\lambda\tau$ and 
\begin{equation}
\omega_{s}=-\frac{(1+s\sqrt{1-c_{z}^{2}})}{2\tau}\left(1-\frac{\lambda_{\text{sv}}^{2}}{2\lambda^{2}}\frac{c_{z}^{2}}{c_{z}^{2}-1+s\sqrt{1-c_{z}^{2}}}\right)\,.\label{eq:omega_s}
\end{equation}
For the minimal Dirac--Rashba model with $\lambda_{\text{sv}}=0$,
we recover the familiar Dyakonov-Perel relation \citep{Zhang_NJP2012},
resulting in an exponentially decaying solution with spin relaxation
time 
\begin{equation}
\left.\tau_{\perp}\right|_{\lambda_{\text{sv}}=0;\lambda\tau\ll1}=(4\lambda^{2}\tau)^{-1}.\label{eq:DP_limit}
\end{equation}
In the latter regime, the spin polarization is lost due to motional
narrowing, yielding its characteristic dependence on the momentum
scattering time $\tau_{\perp}\propto\tau{}^{-1}$ (see e.g., Refs.\,\citep{Fabian_Acta2007,Wu_2010}).
In the opposite limit of resolved spin-splitting $\lambda\tau\gg1$,
electrons complete full Larmor coherent precession cycles between
scattering events, which induce spin-memory loss (see Fig.\,\ref{fig:Spin-relaxation-in}
and discussion below). In this limit, the spin lifetime is of the
order of the momentum scattering time, similarly to two-dimensional
electron gases with large spin splitting \citep{Gridnev01,Schwab_PRB2006,Liu_PRB2011}.
Combining the two limiting cases, we have 
\begin{equation}
\left.S_{z}(t)\right|_{\lambda\gg\lambda_{\text{sv}}}=\begin{cases}
\exp[-4\,\lambda\,^{2}\tau\,t\,(1-4\lambda_{\text{sv}}^{2}\tau^{2})]\,, & \lambda\tau\ll1\,,\\
e^{-t/2\tau}\cos(2\,\lambda\,t(1+\lambda_{\text{sv}}^{2}/\lambda^{2}))\,, & \lambda\tau\gg1\,.
\end{cases}\label{eq:weakRashba_Sz_intra}
\end{equation}
For dominant spin--valley SOC ($\lambda_{\text{sv}}\gg\lambda$),
we find instead 
\begin{align}
\left.S_{z}(t)\right|_{\lambda\ll\lambda_{\text{sv}}} & =\exp\left[-\frac{4\lambda^{2}\tau\,t}{1+4\lambda_{\text{sv}}^{2}\tau^{2}}\right]\,,
\end{align}
which provides the asymptotic behaviour 
\begin{equation}
\left.S_{z}(t)\right|_{\lambda\ll\lambda_{\text{sv}}}=\begin{cases}
\exp[-4\,\lambda\,^{2}\tau\,t(1-4\lambda_{\text{sv}}^{2}\tau^{2})]\,, & \lambda_{\text{sv}}\tau\ll1\,,\\
e^{-\frac{t}{\tau}\frac{\lambda^{2}}{\lambda_{\text{sv}}^{2}}}\,, & \lambda_{\text{sv}}\tau\gg1\,.
\end{cases}\label{eq:weakSV_Sz_intra}
\end{equation}
For the in-plane component a similar procedure leads to
\begin{align}
\left.S_{x}(t)\right|_{\lambda\gg\lambda_{\text{sv}}} & =\begin{cases}
\exp[-2\,\lambda\,^{2}\tau\,t(1-4\lambda_{\text{sv}}^{2}\tau^{2})]\,, & \lambda\tau\ll1\,,\\
\cos\left(\sqrt{2}\lambda t\right)\cos\left(\lambda_{\text{sv}}t\right)e^{-\frac{t}{2\tau}}\,, & \lambda\tau\gg1\,.
\end{cases}\label{eq:Sx_Rashba}
\end{align}
and 
\begin{align}
\left.S_{x}(t)\right|_{\lambda\ll\lambda_{\text{sv}}} & =\begin{cases}
\exp[-2\,\lambda\,^{2}\tau\,t(1-4\lambda_{\text{sv}}^{2}\tau^{2})]\,, & \lambda_{\text{sv}}\tau\ll1\,,\\
\cos\left[2\lambda_{\text{sv}}t\left(1+\frac{\lambda^{2}}{\lambda_{\text{sv}}^{2}}\right)\right]e^{-\frac{t}{2\tau}\frac{\lambda^{2}}{\lambda_{\text{sv}}^{2}}}\,, & \lambda_{\text{sv}}\tau\gg1\,.
\end{cases}
\end{align}
Interestingly, the two weak SOC limits $\lambda\ll\lambda_{\text{sv}}\ll1/\tau$
and $\lambda_{\text{sv}}\ll\lambda\ll1/\tau$ display the same spin
dynamics. The spin--valley term only provides a small correction
to the Dyakonov-Perel spin-relaxation time. From these results, the
SRTA ratio for pure intravalley disorder is readily obtained
\begin{equation}
\zeta=\begin{cases}
1 & \lambda\tau\gg\lambda_{\text{sv}}\tau,\,1\\
1/2 & \text{all other cases}\,,
\end{cases}
\end{equation}
that is 
\begin{figure}
\centering{}\includegraphics[width=0.8\columnwidth]{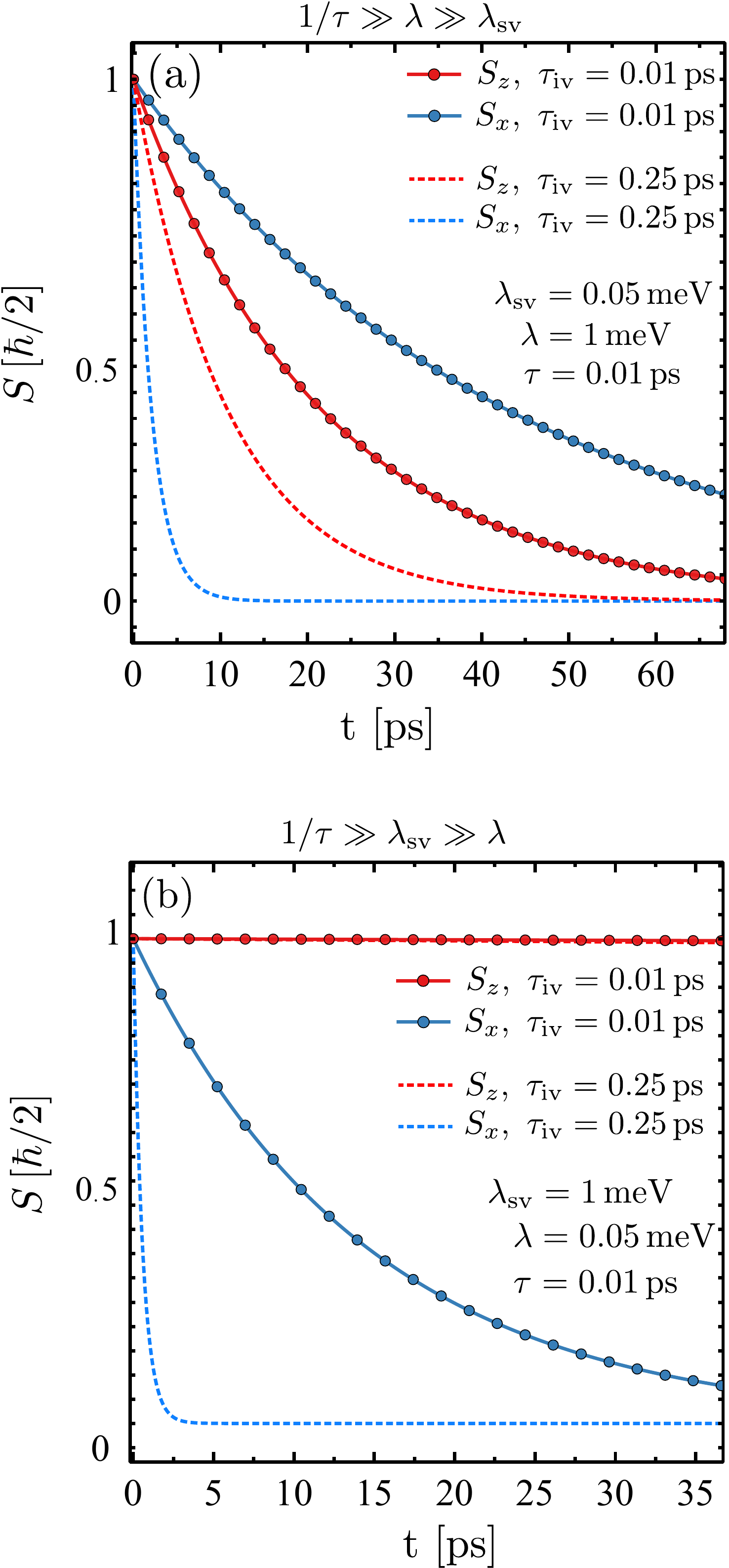}\caption{Spin dynamics for weak proximity-induced SOC in the presence of intervalley
scattering, for $\lambda_{\text{sv}}\ll\lambda$ ($\lambda\ll\lambda_{\text{sv}}$)
{[}panel (a), (b) respectively{]}. The in-plane spin polarization
is more sensitive to the value of $\tau_{\text{iv}}$, whereas out-of-plane
spins are virtually unaffected by a finite $\tau_{\text{iv}}$ in
the limit of very weak Bychkov-Rashba SOC. This is consistent with
the findings in Ref.\,\citep{Cummings_PRL2017}. \label{fig:Spin-dynamics-for-weak}}
\end{figure}
\textbf{ }the conventional SRTA of the minimal ($C_{6v}$) model with
weak Rashba SOC, i.e. $\xi=1/2$ \citep{Fabian_Acta2007} is observed
irrespective of the spin-valley coupling. On the contrary, in the
case of strong Bychkov-Rashba SOC, the quantum kinetic treatment predicts
$\xi=1$. This result is related to the role of the Bychkov-Rashba
field in the two opposite limits $\lambda\tau\ll1$ and $\lambda\tau\gg1$,
cf. Fig.\,\ref{fig:Spin-relaxation-in}. Note that because of the
totally-in plane Bychkov-Rashba SOC, simple commutator algebra for
the precession term $[H_{R},\mathbf{S}]$ gives that the period of
out-of-plane spins is half of that of in-plane ones: $T_{\perp}=T_{\parallel}/2$.
In the Dyakonov-Perel limit $\lambda\tau\ll1$, where electrons' spin
only precesses a small angle before being scattered, the spin dynamics
can be understood as the result of a random walk, with unit step $\delta\varphi_{i}$.
Spin relaxation is achieved after $N$ collisions when the accumulated
phase $\varphi$ is of the orde of unit, that is, $\varphi_{i}\equiv N\,\delta\varphi_{i}\sim1$.
The faster precession of $S_{z}$ reflects in a different unit step
$\delta\varphi_{\perp}=2\delta\varphi_{\parallel}$, which immediately
implies $\tau_{s}^{\perp}=\tau_{s}^{\parallel}/2$, i.e. spins along
$\hat{z}$ reach the critical value $\varphi_{z}\sim1$ in half of
the time compared to initially in-plane spins. On the contrary, when
$\lambda\tau\gg1$, spin relaxation is achieved on the time scale
of a single impurity scattering event, as spins can coherently complete
many precession cycles on a time scale $\tau$. The anisotropic spin
precession reflects instead in this case in the oscillating term rather
than the spin decay, as found in Eqs.\,(\ref{eq:weakRashba_Sz_intra})
and (\ref{eq:Sx_Rashba}).

\subsection{Intervalley scattering case: $w\protect\ne0$}

Short-range scatterers and atomically-sharp defects responsible for
a finite intervalley scattering time $\tau_{\text{iv}}$ are invariably
present in realistic conditions \citep{Wang_PRX2016}. Thus, the inclusion
of intervalley processes in the collision integral is crucial to understand
the spin dynamics in graphene-based heterostructures. Let us start
again from the out-of-plane component. We find in this case 
\begin{equation}
\left(\begin{array}{ccc}
\partial_{t} & -\lambda & 0\\
4\lambda & \partial_{t}-\frac{1}{\tau_{+}} & -2\lambda_{\text{sv}}\\
0 & 2\lambda_{\text{sv}} & \partial_{t}-\frac{1}{\tau_{-}}
\end{array}\right)\left(\begin{array}{c}
S_{z}\\
y\\
z
\end{array}\right)=\left(\begin{array}{c}
0\\
0\\
0
\end{array}\right)\,,\label{eq:Sz0-1}
\end{equation}
with 
\begin{equation}
\frac{1}{\tau_{\pm}}=\frac{1}{\tau}+\frac{2\pm1}{3}\frac{1}{\tau_{\text{iv}}}\,,
\end{equation}
where we have identified the intervalley momentum lifetime\textbf{
} 
\begin{equation}
\tau_{\text{iv}}=\frac{\tau}{3\alpha^{2}}\,.
\end{equation}
Proceeding as shown above, we obtain after standard algebraic manipulations

\begin{equation}
\left.S_{z}(t)\right|_{\lambda\gg\lambda_{\text{sv}}}=\begin{cases}
\exp[-4\,\lambda^{2}\,\tau_{+}\,t\,(1-4\lambda_{\text{sv}}^{2}\tau_{+}\tau_{-})]\,, & \lambda\tau\ll1\,,\\
\cos\left[2\lambda\left(1+\frac{\lambda_{\text{sv}}^{2}}{\lambda^{2}}\right)\right]e^{-t/2\tau_{+}}\,, & \lambda\tau\gg1\,.
\end{cases}
\end{equation}
\begin{equation}
\left.S_{z}(t)\right|_{\lambda\ll\lambda_{\text{sv}}}=\begin{cases}
\exp[-4\,\lambda^{2}\,\tau_{+}\,t(1-4\lambda_{\text{sv}}^{2}\tau_{+}\tau_{-})]\,, & \lambda_{\text{sv}}\tau\ll1\,,\\
e^{-\frac{t}{\tau_{-}}\frac{\lambda^{2}}{\lambda_{\text{sv}}^{2}}}\,, & \lambda_{\text{sv}}\tau\gg1\,.
\end{cases}
\end{equation}

Considering the in-plane components, we were able to reduce the initial
set of 8 coupled equations to two equations coupling $S_{x}=S_{x}^{0}+\bar{S}_{x}^{0}$
and $\tilde{S}_{y}=S_{y}^{0}-\bar{S}_{y}^{0}$ (see Appendix for details),
reading as 
\begin{align}
\left(\begin{array}{cc}
\partial_{t}^{2}+2\lambda^{2}+\frac{1}{\tau_{+}} & 2\lambda_{\text{sv}}(\partial_{t}+\frac{1}{\tau_{+}})\\
-2\lambda_{\text{sv}}(\partial_{t}+\frac{1}{\tau_{*}}) & \partial_{t}^{2}+2\tilde{\lambda}^{2}+\frac{1}{\tau}+\frac{5}{3}\frac{1}{\tau_{\text{iv}}}
\end{array}\right)\left(\begin{array}{c}
S_{x}\\
\tilde{S}_{y}
\end{array}\right)= & \left(\begin{array}{c}
0\\
0
\end{array}\right)\,,\label{eq:Sx_Inter}
\end{align}
where we have set 
\begin{align}
2\tilde{\lambda}^{2} & =2\lambda^{2}+\frac{2}{3}\frac{1}{\tau_{\text{iv}}}\frac{1}{\tau_{*}}\,,\\
\frac{1}{\tau_{*}} & =\frac{1}{\tau}+\frac{1}{3}\frac{1}{\tau_{\text{iv}}}\,.
\end{align}
Solving Eq.\,(\ref{eq:Sx_Inter}) with the same boundary conditions
as above, i.e. $S_{x}(t=0)=1$ and all the other functions being zero
at the initial time, we find 
\begin{figure}
\begin{centering}
\includegraphics[width=1\columnwidth]{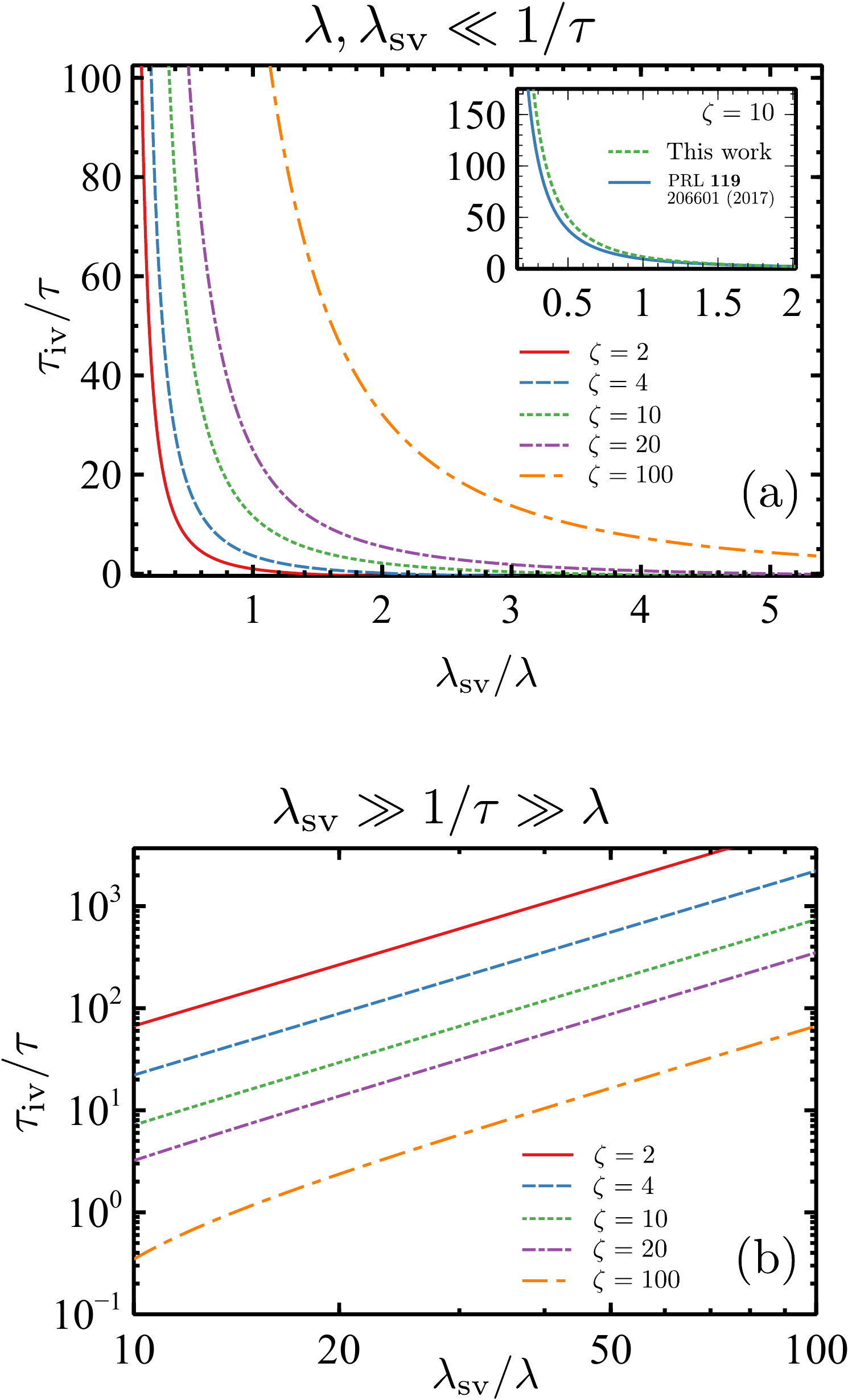}
\par\end{centering}
\caption{Traces of fixed SRTA in the weak (a) and strong (b) proximity-induced
SOC regime, Eq.\,(\ref{eq:SRTA_total}). (a) The inset shows a comparison
with the spin white-noise model in Ref.\,\citep{Cummings_PRL2017}
for $\zeta=10$. (b) Traces of fixed SRTA in the strong proximity-induced
SOC regime, Eq.\,(\ref{eq:SRTA_total}), third line. A typical ratio
$\tau_{\text{iv}}/\tau\sim25$ is compatible with a broad range of
SOCs in the interval $\lambda_{\text{sv}}/\lambda=14-80$ depending
on the measured SRTA ratio. \label{fig:SRTA_traces}}
\end{figure}

\begin{align}
\left.S_{x}(t)\right|_{\lambda\gg\lambda_{\text{sv}}} & =\begin{cases}
\exp[-(2\lambda^{2}\tau_{+}+3\lambda_{\text{sv}}^{2}\tau_{\text{iv}})t]\,,\,\lambda\tau\ll1\,,\\
\\
\exp[-(1/2\tau_{+}+3\lambda_{\text{sv}}^{2}\tau_{\text{iv}})t]\times\\
\times\cos(\sqrt{2}\lambda t)\,,\hfill\lambda\tau\gg1\,,
\end{cases}
\end{align}
and 
\begin{align}
\left.S_{x}(t)\right|_{\lambda\ll\lambda_{\text{sv}}}= & \begin{cases}
\exp[-(2\lambda^{2}\tau_{+}+3\lambda_{\text{sv}}^{2}\tau_{\text{iv}})t]\,,\,\lambda_{\text{sv}}\tau\gg1\,,\\
\\
A\,\cos\left(2\lambda_{\text{sv}}t\right)e^{-\left(\frac{\lambda^{2}}{2\lambda_{\text{sv}}^{2}}\frac{1}{\tau_{+}}+\frac{2}{3\tau_{\text{iv}}}\right)t},\,\lambda_{\text{sv}}\tau\gg1\,.
\end{cases}\label{eq:Sx_full}
\end{align}
with $A=\frac{1+3\tau_{\text{iv}}/\tau}{5+3\tau_{\text{iv}}/\tau}$.
In the large $\lambda_{\text{sv}}$ limit, second line of the latter
equation, the solution includes a second term $B\,e^{-(\tau^{-1}+5\tau_{\text{iv}}^{-1}/3)t}$
with $B=\frac{4}{5+3\tau_{\text{iv}}/\tau}$, giving overall a multi-exponential
solution. This term is subleading in the cases we are interested in,
hence we neglected it in Eq.\,(\ref{eq:Sx_full}). In Figs.\,\ref{fig:Spin-dynamics-for}
and (\ref{fig:Spin-dynamics-for-weak}), we show representative examples
of the spin polarization dynamics in the strong and weak SOC limits,
respectively, according to our results.

\section{Spin relaxation Time Anisotropy }

We discuss now in more detail how the spin dynamics evolves from weak
proximity-induced SOC ($\lambda_{\textrm{SOC}}\tau\ll1$) to well-resolved
SOC ($\lambda_{\textrm{SOC}}\tau\gg1$). The explicit form of the
SRTA ratio is 
\begin{equation}
\zeta=\begin{cases}
\frac{1}{2}+\frac{3}{4}\frac{\lambda_{\text{sv}}^{2}}{\lambda^{2}}\left(1+\frac{\tau_{\text{iv}}}{\tau}\right)\:,\quad\lambda\tau,\lambda_{\text{sv}}\tau\ll1\,,\\
\\
1+O(\lambda_{\text{sv}}^{2}/\lambda^{2})\:,\quad\lambda\tau\gg1\gg\lambda_{\text{sv}}\tau\,,\\
\\
\frac{1}{1+\frac{3\tau_{\text{iv}}}{\tau}}\left[\frac{2\lambda_{\text{sv}}^{2}}{\lambda^{2}}+3\left(1+\frac{\tau_{\text{iv}}}{\tau}\right)\right]\:,\quad\lambda_{\text{sv}}\tau\gg1\gg\lambda\tau\,.
\end{cases}\label{eq:SRTA_total}
\end{equation}
Together with the microscopic derivation of the spin Bloch equation
for this model, Eqs.\,\eqref{eq:system_11}-\eqref{eq:system_23}
and their solution---showing a a crossover between a purely damped
to oscillating damped spin dynamics---these are the most important
results of this paper. The first observation concerns the strong Bychkov-Rashba
case with $\lambda\tau\gg1\gg\lambda_{\text{sv}}\tau$, which can
in principle be achieved in clean graphene-based heterostructures,
where also the lattice mismatch is sizable enough to produce $\lambda_{\text{sv}}\approx0$.\textbf{
}Contrary to the other two presented cases (first and third lines
of Eq.\,(\ref{eq:SRTA_total})), in this limit a direct estimation
of $\tau_{\text{iv}}/\tau$ or $\lambda_{\text{sv}}/\lambda$ is not
possible. Hence, whenever $\zeta\approx1$ is measured, the extraction
of other parameters from spin precession measurements alone should
be considered unfeasible.

We focus in the following on the two more interesting cases $\lambda\tau,\lambda_{\text{sv}}\tau\ll1$
and $\lambda_{\text{sv}}\tau\gg1\gg\lambda\tau.$ For the weak SOC
case---first line of Eq.\,(\ref{eq:SRTA_total})---we report a
visualization of the obtained result in terms of contour lines for
fixed $\zeta$, see Fig.\,(\ref{fig:SRTA_traces})(a). Our results
agrees very well with the toy model supporting the numerical findings
in Ref. \citep{Cummings_PRL2017}, i.e. 
\begin{align}
\zeta & =\frac{1}{2}+\frac{\lambda_{\text{sv}}^{2}}{\lambda^{2}}\frac{\tau_{\text{iv}}}{\tau}\,,\,\,\,\,\,\,\,(\textrm{Ref. 25})\,,\label{eq:zeta_Cummings}\\
\zeta & =\frac{1}{2}+\frac{3}{4}\frac{\lambda_{\text{sv}}^{2}}{\lambda^{2}}\left(1+\frac{\tau_{\text{iv}}}{\tau}\right)\,.\,\,\,\,\,(\textrm{this work})
\end{align}
Note the the different pre-factors in front of the second term with
respect to the results obtained from the microscopic Hamiltonian,
Eq.\,(\ref{eq:SRTA_total})), first line. The inset of Fig.\,\ref{fig:SRTA_traces}(a)
shows a detailed comparison for the case $\zeta=10$. Following the
analysis performed in Ref.\,\citep{Ghiasi_NanoLett2017} ($\zeta=11$),
assuming $\lambda_{\text{sv}}/\lambda\sim0.67$ for graphene/MoSe$_{2}$
\citep{Gmitra_PRB2016}, a $\tau_{\text{iv}}/\tau=30$ is obtained,
which taking $\tau=0.076$ ps gives $\tau_{\text{iv}}=2.2$ ps (against
$\tau_{\text{iv}}=1.7$ ps following Ref.\,\ref{eq:zeta_Cummings}).
These estimtes (obtained from modeling of spin precession data for
$\zeta$) agree qualitatively well with typical relaxation times obtained
from weak localization data \citep{Wang_PRX2016,Tikhonenko_PRL2008}. 

However different scenarios are possible. For instance, in Ref.\,\citep{Yang_2DMat2016},
the authors estimate $\lambda_{\text{sv}}=0.96\,\text{meV}\sim32\,\lambda$,
with $\tau\sim12\,\text{ps}$ for graphene/WS$_{2}$ heterostructures.
In this case the weak SOC approximation might fail. In fact, assuming
$\zeta=11$ as above, using Eq.\,(\ref{eq:zeta_Cummings}) from Ref.\,\citep{Cummings_PRL2017}
one would get an unphysical result $\tau_{\text{iv}}/\tau=0.01<1$,
where the intervalley scattering time is shorter than the (intravalley)
momentum scattering time. The usage of Eq.\,(\ref{eq:SRTA_total})
in the limit of strong spin-valley (third line) then is needed. Using
this relation, we estimate $\tau_{\text{iv}}/\tau\approx70$, pointing
to dominant intravalley processes.

\section{Conclusions}

In this work, we investigated theoretically the spin dynamics in graphene
with proximity-induced SOC. Starting from the quantum Liouville equation,
we derived the effective spin Bloch equations governing the spin dynamics
of 2D Dirac fermions subject to in-plane (Bychkov-Rashba) and out-of-plane
(spin-valley) interactions. We discussed in detail the irreversible
loss of spin information with origin in intra- and inter-valley scattering
processes within the standard Gaussian approximation for the disorder
potential, obtaining the time dependence of the spin polarization
vector and associated spin-relaxation times. We finally discussed
the interesting results for the spin relaxation-time anisotropy $\tau_{s}^{\perp}/\tau_{s}^{\parallel}$.
The result reported Ref.\,\citep{Cummings_PRL2017} for weak SOC
is qualitatively reproduced by our microscopic theory. Crucially,
we have shown that the weak SOC approximation to the spin relaxation
anisotropy ratios fails in when the proximity-induced SOC is of the
same order or larger than the disorder-induced quasiparticle broadening.
Our results for well-resolved SOC then should be used to fit spin
precession measurements.

We remark that the adopted formalism is only valid in the highly-doped
regime of large Fermi energy, where it is assumed that SOC only induces
Larmor precession. This is a strong assumption that might break down
at low electronic density in samples with large interface-induced
SOC of order $1-10$ meV. In that case the spin texture of the bands
is well established; momentum is then strongly correlated with the
direction of the psuedomagnetic field, which can favour or inhibit
certain matrix elements of the scattering potential $U$. For instance,
intervalley scattering has been suggested detrimental for the out-of-plane
spin component, as producing transitions between states with opposite
Zeeman pseudomagnetic field $H_{\text{sv}}^{\kappa}=-H_{\text{sv}}^{\bar{\kappa}}$
\citep{Cummings_PRL2017}. However our treatment, where the kinetic
equations are projected onto eigenstates of bare graphene, has not
the capability of capturing such an effect (see Figs.\,\ref{fig:Spin-dynamics-for}(b),
\ref{fig:Spin-dynamics-for-weak}(b), where $S_{z}$ is virtually
unaffected by the value of $\tau_{\text{iv}}$). A possibility to
incorporate the SOC self-consistently is by adopting the quantum diagrammatic
formalism for Dirac fermions, according to the procedure outlined
in Ref.\,\citep{Offidani_MDPI2018}. We will address this problem
in a future publication. 

\section*{APPENDIX A: DETAILS ON THE DERIVATION OF THE SPIN BLOCH EQUATIONS}

In this appendix we report more details about the derivation of the
spin Bloch equations {[}Eqs.\,\eqref{eq:system_11}-\eqref{eq:system_23}{]}
starting from the collision integral in Eq.\,\eqref{eq:Spin_Bloch}\begin{widetext}
\begin{align}
\left.\partial_{t}\mathbf{S}_{\text{\textbf{k}}}^{\kappa}\right|_{\text{scatt}} & =\langle\mathbf{k}\kappa|I|\mathbf{k}\kappa\rangle\equiv I[\mathbf{S}_{\mathbf{k}}^{\kappa}]\label{eq:S_app_coll}\\
 & =-\pi\sum_{\mathbf{k^{\prime}}\kappa^{\prime}}\delta(\epsilon_{\mathbf{k}}-\epsilon_{\mathbf{k}^{\prime}})\,\langle\mathbf{S}_{\mathbf{k}}^{\kappa}\,U_{\mathbf{kk^{\prime}}}^{\kappa\kappa^{\prime}}\,U_{\mathbf{k}^{\prime}\mathbf{k}}^{\kappa^{\prime}\kappa}+U_{\mathbf{kk^{\prime}}}^{\kappa\kappa^{\prime}}\,U_{\mathbf{k}^{\prime}\mathbf{k}}^{\kappa^{\prime}\kappa}\,\mathbf{S}_{\mathbf{k}}-2\,U_{\mathbf{kk^{\prime}}}^{\kappa\kappa^{\prime}}\,\mathbf{S}_{\mathbf{k^{\prime}}}^{\kappa^{\prime}}\,U_{\mathbf{k}^{\prime}\mathbf{k}}^{\kappa^{\prime}\kappa}\rangle_{\text{dis}}\,
\end{align}
Note the collision integral in Eq.\,\eqref{eq:S_app_coll} is diagonal
in valley space, i.e. $\langle\mathbf{k}\kappa|I|\mathbf{k}\bar{\kappa}\rangle=0$
which was justified in the main text. Intervalley processes are still
taken into account \emph{internally} to the collision integral, i.e.
by considering transitions of the type $K\to K^{\prime}\to K$ where
electrons initially at $K(K^{\prime})$ are scattered at $K^{\prime}(K)$
and then scattered back at $K(K^{\prime})$. 

For point-like impurities, the different matrix elements of the scattering
potential are written as 
\begin{align}
U_{\mathbf{kk^{\prime}}}^{\kappa\kappa^{\prime}} & =\sum_{i=1}^{N}e^{\imath(\mathbf{k^{\prime}-k})\mathbf{x}_{i}}(u_{i}\,\delta_{\kappa\kappa^{\prime}}\cos\phi+\imath\,w_{i}\,\delta_{\kappa\bar{\kappa}\prime}\sin\phi)\,,\\
\phi & \equiv\frac{\phi_{\mathbf{k}^{\prime}}-\phi_{\mathbf{k}}}{2},
\end{align}
which plugged into Eq.\,\eqref{eq:S_app_coll} and after having taken
after disorder average as prescribed in Eqs.\,\eqref{eq:u_1},\eqref{eq:u_2}
gives Eqs.\,\eqref{eq:I_intra}-\eqref{eq:I_inter} of the main text.
Using the notation in the main text and the relation
\begin{equation}
\pi\,n_{i}u^{2}\int_{0}^{\infty}\frac{dk^{\prime}}{2\pi}k^{\prime}\delta(\epsilon_{\mathbf{k}}-\epsilon_{\mathbf{k^{\prime}}})=\frac{n_{i}u^{2}\epsilon}{v^{2}}\equiv\frac{4}{\tau}\,,\label{eq:tau_def}
\end{equation}
 we have explicitly

\begin{align}
I^{\text{intra}}= & -\frac{4}{\tau}\sum_{m}e^{\imath\,m\phi_{\mathbf{k}}}\int_{0}^{2\pi}\frac{d\phi_{\mathbf{k}^{\prime}}}{2\pi}\cos^{2}\left(\frac{\phi_{\mathbf{k}}-\phi_{\mathbf{k}^{\prime}}}{2}\right)\left[1-e^{-\imath\,2m\,\left(\frac{\phi_{\mathbf{k}}-\phi_{\mathbf{k}^{\prime}}}{2}\right)}\right]S_{i}^{m}\\
= & -\frac{4}{\tau}\sum_{m}e^{\imath\,m\,\phi_{\mathbf{k}}}S_{i}^{m}\int_{0}^{2\pi}\frac{d\phi}{2\pi}\cos^{2}\phi\left(1-\cos2\,m\,\phi\right)\\
\equiv & -\sum_{m}e^{\imath\,m\,\phi_{\mathbf{k}}}\frac{S_{i}^{m}}{\tau_{m}^{A}}\,,
\end{align}
with
\begin{equation}
\frac{1}{\tau_{m}^{A}}=\frac{4}{\tau}\int_{0}^{2\pi}\frac{d\phi}{2\pi}\cos^{2}\phi\left[1-\cos(2\,m\,\phi)\right]\,.
\end{equation}
and for the intervalley part 
\begin{align}
I^{\text{inter}}= & -\frac{4}{\tau}\alpha^{2}\left\{ \sum_{m}e^{\imath\,m\,\phi_{\mathbf{k}}}\int_{0}^{2\pi}\frac{d\phi_{\mathbf{k}^{\prime}}}{2\pi}\sin^{2}\left(\frac{\phi_{\mathbf{k}}-\phi_{\mathbf{k}^{\prime}}}{2}\right)\left[S_{i}^{m}-\bar{S}_{i}^{m}e^{-\imath\,2m\,\left(\frac{\phi_{\mathbf{k}}-\phi_{\mathbf{k}^{\prime}}}{2}\right)}\right]\right\} \\
= & -\frac{4}{\tau}\alpha^{2}\left(\sum_{m}e^{\imath\,m\,\phi_{\mathbf{k}}}\frac{S_{i}^{m}}{2}-\bar{S}_{i}^{m}\int_{0}^{2\pi}\frac{d\phi}{2\pi}\sin^{2}\phi\,\cos2\,m\,\phi\right)\\
= & -\alpha^{2}\sum_{m}e^{\imath\,m\,\phi_{\mathbf{k}}}\left(\frac{2S_{i}^{m}}{\tau}-\frac{\bar{S}_{i}^{m}}{\tau_{m}^{B}}\right)\,,
\end{align}
\end{widetext} and
\begin{equation}
\frac{1}{\tau_{m}^{B}}=\frac{4}{\tau}\int_{0}^{2\pi}\frac{d\phi}{2\pi}\sin^{2}\phi\cos(2\,m\,\phi)\,.
\end{equation}
The SOC couples different harmonics $m=0,\pm1,..$.. Let us neglect
that for a while. We find the following system of equations 
\begin{align}
\partial_{t}S_{i}^{m} & =-\left(\frac{1}{\tau_{m}^{A}}+\frac{2\alpha^{2}}{\tau}\right)S_{i}^{m}+\frac{\alpha^{2}}{\tau_{m}^{B}}\bar{S}_{i}^{m}\label{eq:eq:Spin_Valley_Only_01}\\
\partial_{t}\bar{S}_{i}^{m} & =-\left(\frac{1}{\tau_{m}^{A}}+\frac{2\alpha^{2}}{\tau}\right)\bar{S}_{i}^{m}+\frac{\alpha^{2}}{\tau_{m}^{B}}S_{i}^{m}\,,\label{eq:Spin_Valley_Only_0}
\end{align}
and the corresponding expresssion at $K^{\prime}$, obtainable by
$S\to\bar{S}$. Solving them, we have 
\begin{align}
\left(\begin{array}{c}
S_{i}^{m}(t)\\
\bar{S}_{i}^{m}(t)
\end{array}\right) & =e^{-(\frac{1}{\tau_{m}^{A}}+2\alpha^{2}\frac{t}{\tau})}\left(\begin{array}{cc}
\cosh\left(t\frac{\alpha^{2}}{\tau_{m}^{B}}\right) & \sinh\left(t\frac{\alpha^{2}}{\tau_{m}^{B}}\right)\\
\sinh\left(t\frac{\alpha^{2}}{\tau_{m}^{B}}\right) & \cosh\left(t\frac{\alpha^{2}}{\tau_{m}^{B}}\right)
\end{array}\right)\nonumber \\
 & \times\left(\begin{array}{c}
S_{i}^{m}(0)\\
\bar{S}_{i}^{m}(0)
\end{array}\right)\,.
\end{align}
Note for the $m=0$ harmonics, we have $\tau_{0}^{A}\to\infty$ and
$\tau_{0}^{B}\to\tau/2$, so that the solution for the total spin
polarization along $\hat{i}$ is found $S_{i}=S_{i}^{0}+\bar{S}_{i}^{0}=S_{i}(t=0)$,
which is connected to spin conservation in the absence of SOC---the
zeroth-harmonics oscillation of the Fermi surface is associated in
fact with the density. 

Repeating the calculation in the presence of SOC, and considering
$\tau_{\pm1}^{A}=\tau=-\tau_{\pm1}^{B}$ we find Eqs.\,\eqref{eq:system_11}-\eqref{eq:system_23}
of the main text for the leading harmonics $m=0,\pm1$.


\begin{thebibliography}{99}
\bibitem{Foot_1}This is irrelevant for our discussion in this paper
concerning spin relaxation in the highly doped regime, as a term proportional
to $\sigma_{z}$ gives zero when projected onto a single (conduction
or valence) band.
\bibitem{foot_3}\textcolor{black}{Harmonics with $m=\pm1$ contain
information about spin currents, which couple to the spin density
in Eqs.\,\eqref{eq:system_11}-\eqref{eq:system_23} via} the Bychkov-Rashba\textcolor{black}{{}
interaction.}
\end{thebibliography}
\end{document}